\documentclass[trackchanges, preprint2]{aastex701}

\begin{document}

\title{Iskay2: Signal Extraction of the Kinematic Sunyaev-Zel'dovich Effect Through The Pairwise Estimator. Pipeline and Validation.}

\author[orcid=0000-0001-9731-3617,sname='Gallardo']{Patricio A. Gallardo}
\affiliation{Department of Physics and Astronomy, University of Pennsylvania, Philadelphia, PA, USA}
\email[show]{gallardo@upenn.edu}  

\author{Yulin Gong}
\affiliation{Department of Astronomy, Cornell University, Ithaca, NY, USA}
\email{yg527@cornell.edu}

\author{Boryana Hadzhiyska}
\email{boryanah@ast.cam.ac.uk}
\affiliation{Institute of Astronomy, Madingley Road, Cambridge, CB3 0HA, UK}
\affiliation{Kavli Institute for Cosmology Cambridge, Madingley Road, Cambridge, CB3 0HA, UK}

\author{Yun-Hsin Hsu}
\affiliation{University Observatory, Faculty of Physics, Ludwig-Maximilians-Universit\"at, Scheinerstr. 1, 81679 Munich, Germany}
\affiliation{Max-Planck-Institut f\"{u}r extraterrestrische Physik (MPE), Giessenbachstrasse 1, 85748 Garching bei M\"{u}nchen, Germany}
\email{y.hsu@physik.lmu.de}

\begin{abstract}

The peculiar motions of massive halos probe the distribution of matter in the universe, the gravitational potential, and the history of cosmic structure growth. The kinematic Sunyaev-Zeldovich (kSZ) effect offers a robust observational window into these properties. The pairwise kSZ estimator probes the pairwise momentum of groups of galaxies by cross-correlating cosmic microwave background (CMB) maps with spectroscopic galaxy catalogs, using galaxies to trace the positions of dark matter halos. This note introduces iskay2, an efficient pipeline designed to apply the pairwise kSZ estimator to maps of the CMB and large galaxy catalogs. Pairwise kSZ measurements obtained using this pipeline are compared to previously published results and are shown to be consistent within statistical expectations. This pipeline will enable high-precision measurements of the pairwise kSZ utilizing  galaxy catalogs like DESI combined with past, current and next-generation high-resolution CMB experiments such as ACT, SPT and the Simons Observatory.

\end{abstract}


\section{Introduction}

The history of the cosmic expansion of the universe and the properties of dark energy continue to be central topics in modern physics. Observational probes, such as the anisotropies of the microwave background radiation, baryon acoustic oscillations, and type Ia supernovae, have served as powerful tools for constraining the standard $\Lambda\rm CDM$ model. This model describes the accelerating expansion by assuming that the gravitational interactions are mediated by general relativity, while introducing a cosmological constant $\Lambda$, which serves as a dominant dark energy term in the late universe. Possible deviations to the standard model of cosmology would leave imprints in the positions and velocities of massive objects in the universe. Kinematic probes of the distribution of matter provide an opportunity to test for extensions of the standard model of cosmology by probing the gravitational potential and the history of cosmic growth.

\begin{figure*}[t]
    \centering
    \includegraphics[width=0.9\textwidth]{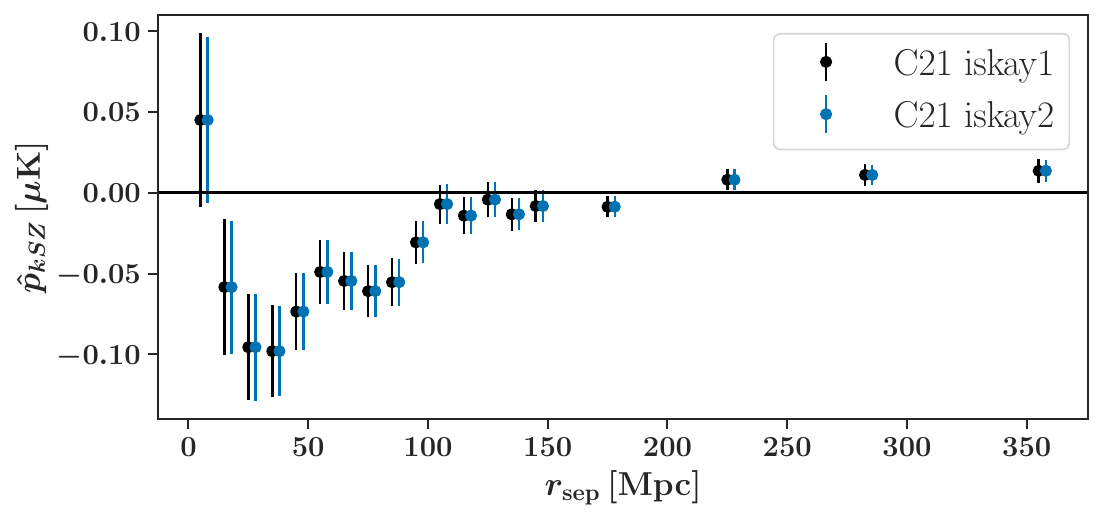}
    \label{iskay2run}
    \caption{Measured pairwise estimator using the full iskay2 pipeline on the ACT DR5 maps compared with published result in C21 showing agreement among the two pipelines. Differences among the mean values are lower than one sigma, and differences among the size of the bootstrapped error bars are consistent with random variance among the two runs. Iskay2 performs $\sim 50\times$ faster than Iskay1 (C21), enabling computation of bootstrapped pairwise kSZ curves using galaxy tracer catalogs of order 1 million tracers, such as DESI.}
\end{figure*}

As it travels through the universe, light from the cosmic microwave background (CMB) is spectrally distorted when it interacts with the hot gas surrounding galaxy clusters (\cite{1972CoASP...4..173S}). In addition to this thermal distortion, the peculiar velocities of galaxy clusters generate a Doppler-shift in the SZ signal known as the kinematic SZ effect (kSZ) (\cite{1980MNRAS.190..413S}). The kSZ is an order of magnitude smaller than the thermal SZ effect in amplitude and is challenging to extract from CMB data. However, statistical detections of the kSZ have been possible with the introduction of wide and sensitive maps of the CMB, cross-correlated with galaxy catalogs as in: \cite{PhysRevLett.109.041101,Soergel10.1093/mnras/stw1455,Bernardis_2017,Schiappucci_PhysRevD.107.042004}, and \cite{Calafut21_PhysRevD.104.043502} (hereafter C21). 

The increasing data volumes introduced by current surveys pose the computational challenge of efficient and validated pipelines. This document presents iskay2, a pipeline that reproduces previously published pairwise kSZ measurements while providing  efficient computation times. This pipeline will enable future high-significance detections using large galaxy catalogs consisting of over 1 million galaxies, such as DESI.

\section{The Pairwise kSZ estimator}

The pairwise kSZ estimator makes use of a large spectroscopic galaxy catalog and a sensitive, high-resolution map of the CMB. The spectroscopic galaxy catalog is used to obtain the three-dimensional directions within pairs of galaxy haloes as a function of separation, and using the direction of each pair $c_{ij}$ as a weight in the maximum-likelihood estimator as in \cite{Ferreira1999} \begin{equation}
    \label{eq:ferreiraestimator}
    \hat p(r) = -\frac{\sum c_{ij} dT_{ij}}{\sum c_{ij}^2},
\end{equation}
where $c_{ij} = \frac12 \hat r_{ij} \cdot ({\hat r_i - \hat r_j})$ is the projection of the direction joining the pair and the line of sight, and $dT_{ij} = T_{AP}^i - T_{AP}^j$ is the difference of the intensity for the pair obtained with an aperture photometry filter and corrected by a possible redshift dependence as in C21.
The sums in Eq. \ref{eq:ferreiraestimator} must be carried out for a range of cosmological separations. In addition, the bootstrapped estimation of error bars and covariance matrices for this estimator require repeated computation of this sum. This presents a computational challenge as a naive implementation of this estimator takes an unrealistic time to compute for a catalog that exceeds $\sim 10^6$ objects, like in modern and next-generation spectroscopic surveys.

\section{Iskay2}

Iskay2 is a Python library which makes use of the set of high-performance routines to measure clustering statistics Corrfunc (\cite{Corrfunc.2020MNRAS.491.3022S}), which have been adapted to compute the pairwise estimator and give support for variants such as weighted averaged versions of the same estimator. Code is publicly available (\cite{patogallardo_2025_17364285}).

\section{Result, Validation and Conclusion}

Figure \ref{iskay2run} shows consistency among the pairwise estimator applied to the ACT dataset in conjunction with the SDSS catalog of 343k LRGs as discussed in C21. Mean values deviate less than one sigma, which validates the pipeline. The size of the error bar is consistent among different pipeline runs, except for random variations due to the random resampling. Iskay2 provides a $\sim 50\times$ improvement in speed, allowing the computation of a pairwise kSZ curve with a bootstrapped covariance matrix for 1 million tracers in under an hour runtime.

New maps of the CMB such as the Atacama Cosmology Telescope DR6 in conjunction with large spectroscopic catalogs of galaxies such as DESI provide a unique opportunity to detect the pairwise kSZ signal with statistical significance well over 6$\sigma$. In addition, the Simons Observatory will provide even more precise maps with wide coverage, which combined with future data releases from DESI will improve constraints in the kSZ signal by an order of magnitude in the next decade.

\software{Matplotlib (\cite{Hunter:2007}), Numpy (\cite{harris2020array}), Corrfunc (\cite{Corrfunc.2020MNRAS.491.3022S}), Pixell (\cite{pixell.2021ascl.soft02003N}).}

\bibliography{sample701}{}
\bibliographystyle{aasjournalv7}

\end{document}